\documentstyle[twocolumn,prb,aps,epsfig,floats,eqsecnum]{revtex}

\begin{document}
\draft

\twocolumn[\hsize\textwidth\columnwidth\hsize\csname @twocolumnfalse\endcsname

\title{Spin and orbital effects of Cooper pairs coupled to a single magnetic impurity}

\author{Mikito Koga$^1$ and Masashige Matsumoto$^2$}
\address{$^1$Department of Physics, Faculty of Education, Shizuoka University, 836 Oya,
Shizuoka 422-8529, Japan \\
$^2$Department of Physics, Faculty of Science, Shizuoka University, 836 Oya,
Shizuoka 422-8529, Japan}

\date{\today}
\maketitle

\begin{abstract}
The Kondo effect strongly depends on spin and orbital degrees of freedom of
unconventional superconductivity.
We focus on the Kondo effect in uniformly gapped superconducting systems
(the two-dimensional $p_x + i p_y$-wave and $d_{x^2 - y^2} + i d_{xy}$-wave
superconductors here)
to compare the magnetic properties of the spin-triplet and spin-singlet Cooper pairs.
The difference appears when both of the paired electrons couple to a local spin directly.
For the $p_x + i p_y$-wave, the ground state is always a spin doublet for
a $S_{\rm imp} = 1/2$ local spin, and it is always a spin singlet for $S_{\rm imp} = 1$.
The latter is due to uniaxial spin anisotropy of the triplet Cooper pair.
For the $d_{x^2 - y^2} + i d_{xy}$-wave, the interchange of ground states occurs, which
resembles a competition between the Kondo effect and the superconducting energy gap
in $s$-wave superconductors.
Thus the internal degrees of freedom of Cooper pairs give a variety to the Kondo effect.
\end{abstract}

\vspace{10pt}

]\narrowtext

\newcommand{\br}{{\mbox{\boldmath$r$}}}
\newcommand{\bk}{{\mbox{\boldmath$k$}}}
\newcommand{\sk}{{\mbox{\footnotesize $k$}}}
\newcommand{\bsk}{{\mbox{\footnotesize \boldmath$k$}}}
\newcommand{\bS}{{\mbox{\boldmath$S$}}}
\newcommand{\bd}{{\mbox{\boldmath$d$}}}
\newcommand{\bsigma}{{\mbox{\boldmath$\sigma$}}}
\newcommand{\fig}[1]
{\vspace{24pt}
\begin{center}
\fbox{\rule{0cm}{#1}\hspace{7cm}}
\end{center}}

\section{introduction}
The problem of magnetic impurities in superconductors has been investigated for
a long time.
\cite{Shiba68,Sakurai,Muller,Matsuura,Jarrell,Satori,Sakai,Yoshioka}
This Kondo problem was solved at least for the case of a standard BCS ($s$-wave)
superconductor.
Since the superconducting energy gap $\Delta$ competes with the Kondo temperature
$T_{\rm K}$, the Cooper pairs are partly destroyed and a local spin tries to
couple with the conduction electrons antiferromagnetically to form a local singlet
(Kondo singlet).
As a consequence, the Kondo singlet is stabilized for $T_{\rm K} / \Delta \gg 1$,
while the local spin is almost free from the Cooper pairs for $T_{\rm K} / \Delta \ll 1$.
This shows that a singlet ground state can be realized for large $T_{\rm K}$ even
if the energy gap exists.
An interesting question arises in mind when we introduce spin and orbital degrees
of freedom into Cooper pairs:
is the Kondo singlet still stable for large values of $T_{\rm K} / \Delta$
in unconventional superconductors?
In general, the unconventional superconductivity is characterized by a spin and
momentum dependent order parameter $\hat{\Delta}_{\sigma \sigma'}(\bk)$.
The $\bk$ dependence is classified into several angular momenta ($s$, $p$, $d$, $f$) or
these combinations.
A gapless superconductor expressed by $\Delta_{\uparrow \downarrow}(k_x^2 - k_y^2)$
is observed in the high-$T_{\rm c}$ cuprates.
\cite{Tsuei}
A possibility of spin-triplet superconductor
is expected for some heavy fermion compounds and ruthernates:
the simplest expression of the $\bk$ dependence (orbital part) of the order parameter
is given by $(k_x + i k_y)$, which is proposed for time-reversal
breaking superconductors, UGe$_2$
\cite{Saxena,Machi}
and Sr$_2$RuO$_4$.
\cite{Maeno,Sigrist}
In the experimental side, much effort has been made to detect evidence
of the Kondo effect below the superconducting transition temperature in high-$T_{\rm c}$
cuprates.
A recent nuclear magnetic resonance experiment reported that a local magnetic
susceptibility shows reduction of Kondo screening due to opening of the superconducting
energy gap.
\cite{Bobroff}
\par

In this paper, we focus on two simple unconventional superconducting states which have
a full energy gap and whose angular momenta are good quantum numbers, assuming that
a local spin is located at the center of two-dimensional coordinates.
One is a $p_x + i p_y$-wave state whose order parameter is given by
$\Delta_{\uparrow \downarrow}(k_x + i k_y)$, and the other is
a $d_{x^2 - y^2} + i d_{xy}$-wave one represented by
$\Delta_{\uparrow \downarrow}([k_{x^2} - k_{y^2}] + i k_x k_y)$.
Since the density of quasiparticle states for both cases has the same energy gap with
the $s$-wave, we can extract an orbital effect of Cooper pairs which is never
seen in the $s$-wave superconductivity.
We can clarify a spin effect in the Kondo effect as well to compare the spin-triplet
pair with the spin-singlet one.
We investigate a new type of Kondo effect due to a $S_{\rm imp} = 1/2$
or $S_{\rm imp} = 1$
impurity ($S_{\rm imp}$ is a size of the local spin) in these superconducting states,
using the numerical renormalization group (NRG) method.
\cite{Wilson,Sakai92}
In our NRG model, it is found that
the quasiparticles of the superconducting states have two channels which are coupled
to each other only at the impurity.
The NRG technique is very powerful to study such complicated Kondo problems.
\par

In our previous study on the Kondo effect in the unconventional superconductors,
\cite{Matsumoto01,Matsumoto01-2}
a short-range scattering object was considered for the impurity.
The atomic orbital of the impurity is restricted to the $s$-orbital denoted by
the $l = 0$ angular momentum.
The angular momentum of the Cooper pair is $l = 1$ for the $p_x + i p_y$-wave and
$l = 2$ for the $d_{x^2 - y^2} + i d_{xy}$-wave.
In metallic states, $l = 0$ is the only relevant orbital in the Kondo effect.
For the $p_x + i p_y$-wave, the Cooper pair consists of $l = 0$ and $l =1$ electrons
and involves the $l = 1$ electrons in the Kondo effect.
In the same manner, $l = 2$ electrons become relevant for the
$d_{x^2 - y^2} + i d_{xy}$-wave.
This leads to a doublet ground state and the situation is completely different from
the $s$-wave superconducting case.
However, in this case, we did not find any difference between the $p_x + i p_y$-wave and
$d_{x^2 - y^2} + id_{xy}$-wave cases.
\par

In the present study, we take into account the scatterings of the $l = 1$ ($l = 2$)
electrons on the impurity site as well as $l = 0$ for the $p_x + i p_y$-wave
($d_{x^2 - y^2} + i d_{xy}$-wave).
Introducing the two exchange couplings, we can distinguish the spin-triplet Cooper
pair from the spin-singlet one in the Kondo effect.
The difference appears in the form of the effective exchange interaction in
our Kondo model.
The $p_x + i p_y$-wave generates spin anisotropy, and the ground state is always
a spin doublet for $S_{\rm imp} = 1/2$.
For $S_{\rm imp} = 1$, the triplet state of the local spin is split into a singlet and
a doublet, and the ground state is always a spin singlet.
For the $d_{x^2 - y^2} + i d_{xy}$-wave, the ground state changes from a spin doublet
to a particle-hole doublet as $T_{\rm k}$ increases for $S_{\rm imp} = 1/2$.
Similarly, the interchange of triplet and singlet ground states occurs
for $S_{\rm imp} = 1$.
\par

This paper is organized as follows.
The NRG formulation of our Kondo problem is presented in Sec.~II.
In Sec.~III, the difference between the two types of unconventional superconductivity is
discussed, based on the NRG results.
The paper is closed with concluding remarks in Sec.~IV.
\par
\section{model}
First we summarize the derivation of the NRG Hamiltonian for the Kondo effect
in the $p_x + i p_y$-wave ($d_{x^2 - y^2} + i d_{xy}$-wave) superconductor.
\cite{Matsumoto01-2}
One can find that this type of superconducting order parameter is uniformly
gapped since $\Delta(k_x + i k_y)$ for the $p_x + i p_y$-wave is expressed
with $\Delta e^{i \phi_\sk}$,
where $\phi_\sk$ is the angle of a wave vector $\bk$ measured from the $k_x$ axis
and $\Delta$ is a real function of $|\bk|$.
We note that this $\Delta$ is not a function of $\bk$ itself in this case.
\par

Let us begin with the following Hamiltonian:
\begin{eqnarray}
&& H=H_{\rm kin} + H_\Delta + H_{\rm imp}^s,
\label{eqn:2.1} \\
&& H_{\rm kin} = \sum_{\bsk\sigma} \varepsilon_\bsk
   a_{\bsk\sigma}^\dagger a_{\bsk\sigma}, \\
&& H_\Delta = \sum_\bsk \left( \Delta_\bsk
   a_{\bsk\uparrow}^\dagger a_{-\bsk\downarrow}^\dagger + {\rm H. c.} \right), \\
&& H_{\rm imp}^s = -\frac{J}{2} \sum_{\bsk\bsk'\sigma\sigma'}
   \bS \cdot \bsigma_{\sigma\sigma'} a_{\bsk\sigma}^\dagger a_{\bsk'\sigma'}.
\end{eqnarray}
Here $H_{\rm kin}$, $H_{\Delta}$, and $H_{\rm imp}^s$ represent the kinetic energy of
conduction electrons, the BCS interaction, and the effective exchange interaction at
the impurity, respectively.
The operator $a_{\bsk\sigma}^\dagger$ ($a_{\bsk\sigma}$) represents
creation (annihilation) of a conduction electron with a wave vector $\bk$ and
spin $\sigma$.
In $H_{\rm imp}^s$, $\bsigma$ is the Pauli matrix for the conduction electrons.
The impurity is expressed by the local spin operator $\bS$.
The exchange coupling $J (<0)$ is antiferromagnetic.
Since the orbital angular momentum is a good quantum number
for the $p_x + i p_y$-wave Cooper pairs,
we can simplify the Hamiltonian by expanding $a_{\bsk \sigma}$ with respect to the
two-dimensional polar coordinate bases:
\begin{equation}
a_{\bsk\sigma} = \sum_l (-i)^l |J_{l+1}(kR)| e^{i l \phi_\sk} a_{\sk l\sigma},
\end{equation}
where $J_{l}$ is the $l$-th Bessel function, $R$ is the radius of the two-dimensional
system, and $l$ is the $z$-component of the orbital angular momentum of
the conduction electron.
The Hamiltonian~(\ref{eqn:2.1}) is then expressed as
\begin{eqnarray}
&& H_{\rm kin} = \sum_{k l \sigma}
   \varepsilon_k a_{\sk l \sigma}^\dagger a_{\sk l \sigma}, \\
&& H_\Delta = \sum_{k l} (-1)^{1-l} \left(i\Delta a_{\sk l \uparrow}^\dagger
   a_{\sk -l + 1, \downarrow}^\dagger + {\rm H. c.} \right), \\
&& H_{\rm imp}^s = -\frac{J}{2} \sum_{k k'\sigma \sigma'}
   \bS \cdot \bsigma_{\sigma\sigma'} a_{\sk 0 \sigma}^\dagger a_{\sk' 0 \sigma'}.
\label{eqn:2.8}
\end{eqnarray}
In Eq.~(\ref{eqn:2.8}), we have assumed a short-range scattering impurity.
It is sufficient to choose the $l = 0$ and $l =1$ orbitals and neglect others for the
$p_x + i p_y$-wave case, since they are the only orbitals of conduction electrons that
participate in the Kondo effect.
The $l = 1$ electrons are coupled to the local spin via the superconducting
order parameter.
The most appropriate forms of the three parts of Hamiltonian~(\ref{eqn:2.1}) for the NRG
calculation are finally given by
\begin{eqnarray}
&& H_{\rm kin} = \sum_{k \sigma} \varepsilon_k
   \left( a_{\sk 0 \sigma}^\dagger a_{\sk 0 \sigma}
        + a_{\sk 1 \sigma}^\dagger a_{\sk 1 \sigma} \right), \\
&& H_\Delta = \sum_{k} \left(i\Delta
   a_{\sk 1\sigma}^\dagger a_{\sk 0,-\sigma}^\dagger
                    + {\rm H. c.} \right), \\
&& H_{\rm imp}^s = -\frac{J}{2} \sum_{k k'\sigma \sigma'} \bS \cdot \bsigma_{\sigma\sigma'}
   a_{\sk 0 \sigma}^\dagger a_{\sk' 0 \sigma'}.
\end{eqnarray}
A similar Hamiltonian for the $d_{x^2 - y^2} + i d_{xy}$-wave can be obtained if
$i \Delta$ and $l = 1$ are replaced by $-\sigma \Delta$ and $l = 2$, respectively,
since the order parameter $\Delta([k_{x^2} - k_{y^2}] + i k_x k_y)$ is expressed with
$\Delta e^{i 2 \phi_\sk}$.
\par

Next we present the derivation of the NRG Hamiltonian from the Kondo model.
For simplicity, we use the same values for both the superconducting cutoff and the
band width, since it is assured that this does not alter the results.
\cite{Sakai}
After applying the Wilson's logarithmic discretization procedure to the conduction band,
\cite{Wilson}
we obtain the following hopping type of Hamiltonian:
\begin{eqnarray}
H_{\rm kin} &=& {1 + \Lambda^{-1} \over 2} \cr
&\times& \sum_{n =0}^{\infty} \sum_{l\sigma}
\Lambda^{-n/2} \varepsilon_n (f_{nl\sigma}^{\dag} f_{n + 1,l\sigma}
 + f_{n + 1,l\sigma}^{\dag} f_{nl\sigma}),
\end{eqnarray}
for the free electrons, where $l$ takes $0$ and $1$, and
\begin{eqnarray}
H_{\Delta} = \sum_{n = 0}^{\infty} \sum_{\sigma}
(-i \Delta) (f_{n,0,\sigma}^{\dag} f_{n,1,-\sigma}^{\dag}
 - f_{n,1,-\sigma} f_{n,0,\sigma}),
\end{eqnarray}
for the BCS pairing interaction in the $p_x + i p_y$-wave superconductor.
To diagonalize $H_{\Delta}$, we use the following Bogoliubov transformation:
\begin{eqnarray}
&& b_{n,+,\sigma}^{\dag} = {1 \over \sqrt{2}} (f_{n,0,\sigma}^{\dag} - i
f_{n,1,-\sigma}),
\label{eqn:2.14} \\
&& b_{n,-,-\sigma} = {1 \over \sqrt{2}} (f_{n,1,-\sigma} - i
f_{n,0,\sigma}^{\dag}),
\label{eqn:2.15}
\end{eqnarray}
and we obtain
\begin{eqnarray}
H_{\Delta} = -\Delta \sum_{n = 0}^{\infty}
\left(\sum_{\tau = \pm \sigma} b_{n \tau \sigma}^{\dag} b_{n \tau \sigma} - 2 \right).
\end{eqnarray}
By applying the following particle-hole transformation:
\cite{Satori}
\begin{eqnarray}
&& c_{2n,+,\sigma}^{\dag} = b_{2n,+,\sigma}^{\dag},~~
c_{2n,-,\sigma}^{\dag} = i b_{2n,-,-\sigma}, \nonumber \\
&& c_{2n-1,+,\sigma}^{\dag} = -i b_{2n-1,-,-\sigma},
\nonumber \\
&& c_{2n-1,-,\sigma}^{\dag} = -b_{2n-1,+,\sigma}^{\dag},
\label{eqn:2.17}
\end{eqnarray}
we obtain the standard form of NRG Hamiltonian in a recursion relation
\begin{eqnarray}
&&H_{N+1} = \Lambda^{1/2}H_N + \sum_{\tau\sigma}
  \Bigl[
    \varepsilon_N ( c_{N+1,\tau\sigma}^\dagger c_{N\tau\sigma} + {\rm H.c.} ) \cr
&&~~~~~~~~~~~~~~~
     +(-1)^N\Lambda^{N/2}\tau\tilde{\Delta}
        c_{N+1,\tau\sigma}^\dagger c_{N+1,\tau\sigma}
  \Bigr],
\label{eqn:2.18}
\end{eqnarray}
and
\begin{eqnarray}
&&H_0 =
  \Bigl[ -\frac{\tilde{J}}{2}
    \sum_{\tau\tau'\sigma\sigma'}
      \mbox{\boldmath$S$} \cdot \mbox{\boldmath$\sigma$}_{\sigma\sigma'}
        c_{0\tau\sigma}^\dagger c_{0\tau'\sigma'} \cr
&&~~~~~~~~~~~~~~~ -\sum_{\tau\sigma} \tau\tilde{\Delta}
  c_{0\tau\sigma}^\dagger c_{0\tau\sigma} \Bigr] \Lambda^{-1/2},
\label{eqn:2.19}
\end{eqnarray}
for the impurity.
Here $\tilde{J}$ and $\tilde{\Delta}$ are given by
\begin{eqnarray}
&& \tilde{J} = {2 \over 1 + \Lambda^{-1}} J \rho, \\
&& \tilde{\Delta} = {2 \over 1 + \Lambda^{-1}} \Delta,
\end{eqnarray}
where $\rho$ is the density of states at the Fermi energy and $\Delta$ is
normalized by the conduction band width here.
The above procedure can be applied to the $d_{x^2 - y^2} + i d_{xy}$-wave case,
and the same form of the NRG Hamiltonian is obtained.
\par

In the above argument, we have restricted an orbital type of the scattering impurity
to the s-wave ($l = 0$).
In this case, both $p_x + i p_y$-wave and $d_{x^2 - y^2} + i d_{xy}$-wave superconducting
states exhibit the same Kondo behavior resulting from Eqs.~(\ref{eqn:2.18})
and (\ref{eqn:2.19}).
The simplest way to distinguish the two types of pairings is that
we add $p$-wave scattering ($l = 1$) for the $p_x + i p_y$-wave case and $d$-wave
scattering ($l = 2$) for the $d_{x^2 - y^2} + i d_{xy}$-wave.
First, we discuss an extended Kondo Hamiltonian for the former, which is given by
\begin{equation}
H_{\rm imp}^{s+p} = \sum_{\sigma \sigma'} \bS \cdot \bsigma_{\sigma \sigma'}
(-J_0 f_{00 \sigma}^{\dagger} f_{00 \sigma}
\mbox{} -J_1 f_{01 \sigma}^{\dagger} f_{01 \sigma'}),
\end{equation}
where $J_0$ and $J_1$ are coupling constants for the $l = 0$ and $l = 1$
electrons, respectively, and the fermion operator $f_{0 l \sigma}$ is defined by
\begin{equation}
f_{0 l \sigma} = {1 \over \sqrt{2}} \sum_k a_{k l \sigma}.
\end{equation}
{}From Eqs.~(\ref{eqn:2.14}), (\ref{eqn:2.15}), and (\ref{eqn:2.17}), we obtain
\begin{eqnarray}
&& f_{0 0 \sigma}^{\dagger} = {1 \over \sqrt{2}}
   (c_{0,+,\sigma}^{\dagger} + c_{0,-,\sigma}^{\dagger}), \\
&& f_{0 1, -\sigma} = {i \over \sqrt{2}}
   (c_{0,+,\sigma}^{\dagger} - c_{0,-,\sigma}^{\dagger}),
\label{eqn:2.23}
\end{eqnarray}
and transform $H_{\rm imp}^{s+p}$ to the NRG Hamiltonian for the impurity
\begin{eqnarray}
&&H_0 =
  \Bigl[ -\frac{\tilde{1}}{2}
    \sum_{\tau\tau'} \{
      \tilde{J}_{\tau \tau'}^z S_z (c_{0 \tau \uparrow}^{\dagger} c_{0 \tau' \uparrow}
\mbox{} - c_{0 \tau \downarrow}^{\dagger} c_{0 \tau' \downarrow}) \cr
&&~~~~~~~~~~~~~~~~
\mbox +~\tilde{J}_{\tau \tau'}^\bot (S_- c_{0\tau\uparrow}^\dagger c_{0\tau'\downarrow}
\mbox ~+~ S_+ c_{0\tau\downarrow}^\dagger c_{0\tau'\uparrow}) \} \cr
&&~~~~~~~~~~~~~~~ -\sum_{\tau\sigma} \tau\tilde{\Delta}
  c_{0\tau\sigma}^\dagger c_{0\tau\sigma} \Bigr] \Lambda^{-1/2},
\label{eqn:2.24}
\end{eqnarray}
where $S_{\pm} = S_x \pm i S_y$, and the orbitally dependent exchange couplings are
given by
\begin{eqnarray}
&& {\tilde J}_{\tau \tau'}^z = \tilde{J}_0 + {\rm sgn}(\tau \tau') \tilde{J}_1, \cr
&& {\tilde J}_{\tau \tau'}^{\bot} = \tilde{J}_0 - {\rm sgn}(\tau \tau') \tilde{J}_1.
\end{eqnarray}
We can find that uniaxial anisotropy is introduced to the spin of the total system
through the couplings:
$\tilde{J}_{\tau \tau'}^z \ne \tilde{J}_{\tau \tau'}^{\bot}$ for finite values of
$\tilde{J}_0$ and $\tilde{J}_1$.
On the other hand, for the $d_{x^2 - y^2} + d_{xy}$-wave, the transformation
corresponding to Eq.~(\ref{eqn:2.23}) is given by
\begin{equation}
f_{0 2, -\sigma} = {\sigma \over \sqrt{2}}
(c_{0,+,\sigma}^{\dagger} - c_{0,-,\sigma}^{\dagger}),
\end{equation}
for the $l = 2$ electrons.
The $H_0$ for the $l = 0$ and $l = 2$ electrons is given by
the same form with Eq.~(\ref{eqn:2.24}) except for the exchange couplings
\begin{eqnarray}
&& \tilde{J}_{\tau \tau'}^z = \tilde{J}_0 + {\rm sgn}(\tau \tau') \tilde{J}_2, \cr
&& \tilde{J}_{\tau \tau'}^{\bot} = \tilde{J}_0 + {\rm sgn}(\tau \tau') \tilde{J}_2,
\end{eqnarray}
where $\tilde{J}_2$ is a coupling constant between the impurity and $l = 2$ electrons.
In this case, the total spin keeps its isotropy:
$\tilde{J}_{\tau \tau'}^z = \tilde{J}_{\tau \tau'}^{\bot}$ is satisfied for all values
of $\tilde{J}_0$ and $\tilde{J}_2$.
Thus, by introducing the $l \ne 0$ scattering into the exchange interaction,
we can see the difference between the triplet and singlet pairings
which comes from the phases of the wave functions of the $l \ne 0$ electrons.
\par

\section{results}
\begin{figure}
\begin{center}
\epsfxsize=7.5cm
\epsfbox{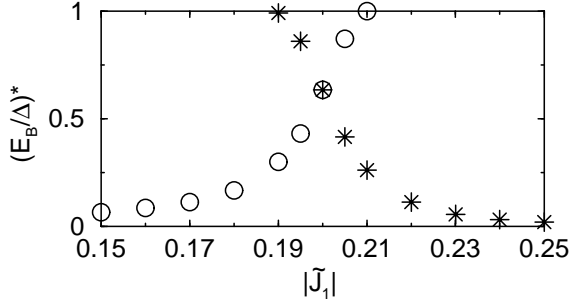}
\end{center}
\caption{
$\tilde{J}_1$ ($<0$) dependence of the bound state energy levels
for $S_{\rm imp} = 1/2$ in the $p_x + i p_y$-wave superconducting state,
where $\tilde{J}_0=-0.2$ and $\tilde{\Delta} = 10^{-5}$ ($\tilde{\Delta} = 1.5 \Delta$).
The energies are measured from those of the spin doublet ground state ($S_z = \pm 1/2$).
The two bound states are particle-hole doublets ($S_z = 0$).
The circles and stars represent data for one bound state dominated by $\tilde{J}_0$
and the other dominated by $\tilde{J}_1$, respectively.
In the NRG calculation, the lowest-lying $\sim$500 states are kept at each renormalization
step.
}
\label{fig:1}
\end{figure}
\subsection{Short-range scattering impurity}
First we summarize main results for a case of short-range scattering impurity with
$S_{\rm imp} = 1/2$, the details of which were discussed in our previous papers.
\cite{Matsumoto01,Matsumoto01-2}
Both $\tilde{J}_0$ and $\tilde{\Delta}$ are relevant in the Kondo effect, growing as
$\Lambda^{N/2}$ with the renormalization step $N$ of the Hamiltonian~(\ref{eqn:2.18}).
The competition between the Kondo effect and the superconductivity is found in
$T_{\rm K} / \Delta$ dependence of the bound state energy.
The ratio of this energy and the renormalized $\tilde{\Delta}$ approaches a constant
value when $\tilde{\Delta}$ becomes large enough.
In a logarithmic scale of $T_{\rm K} / \Delta$, the convergent
value $(E_{\rm B} / \Delta)^*$
measured from a spin-doublet ground state changes monotonically from a unity for
$T_{\rm K} / \Delta \ll 1$ to zero for $T_{\rm K} / \Delta \gg 1$.
This means that the ground state is always a spin doublet for the $p_x + i p_y$-wave
superconductivity.
However, the Kondo effect does occur since $T_{\rm K}$ is found to be an important
energy scale.
In fact, the local spin is shrunk as $T_{\rm K} / \Delta$ increases.
It is shown by the decrease of an effective $g$-factor which is estimated from Curie-law
behavior of the local spin susceptibility $\chi_{\rm loc}$.
\cite{Matsumoto01-2}
In the strong coupling limit ($T_{\rm K} / \Delta \gg 1$), the local spin is almost
quenched by the $l = 0$ electrons, while the $l \ne 0$ electrons couple weakly with
the local spin via the superconducting order parameter to gain the superconducting
condensation energy.
Therefore, a singlet ground state is not realized even if $T_{\rm K}$ is very large,
which is completely different from the $s$-wave superconducting case.
For the latter, the interchange of singlet and doublet ground states occurs.
\par

In the above result, we did not see any difference between the triplet
($p_x + i p_y$-wave) and singlet ($d_{x^2 - y^2} + i d_{xy}$-wave) pairing states.
For both pairings, the orbital effect of the Cooper pairs is important in the Kondo effect,
while their spin does not affect the Kondo effect.
\par

\begin{figure}
\begin{center}
\epsfxsize=7.5cm
\epsfbox{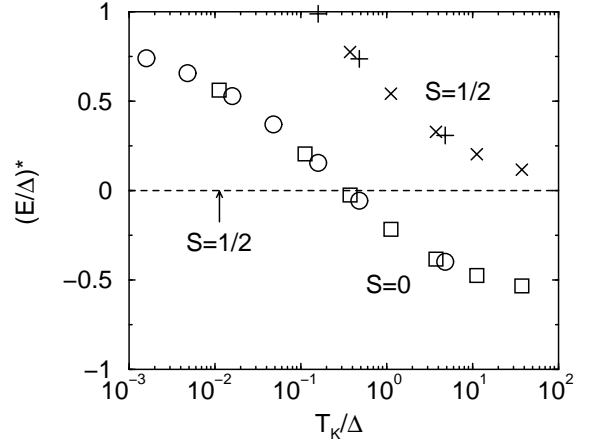}
\end{center}
\caption{
$T_{\rm K} / \Delta$ dependence of the energies of the lowest-lying particle-hole
doublet ($S = 0$) and the second lowest-lying spin doublet ($S = 1/2$)
for $S_{\rm imp} = 1/2$ in the $d_{x^2 - y^2} + i d_{xy}$-wave superconducting state.
The energies are measured from those of the lowest-lying spin doublet ($S = 1/2$).
The circles ($\tilde{J}_0 = \tilde{J}_2 = -0.2$) and squares
($\tilde{J}_0 = \tilde{J}_2 = -0.3$) represent the data for the $S = 0$ state.
The plus ($\tilde{J}_0 = \tilde{J}_2 = -0.2$) and cross symbols
($\tilde{J}_0 = \tilde{J}_2 = -0.3$) represent the data for the second
lowest-lying $S = 1/2$ state.
The interchange of the $S = 1/2$ and $S = 0$ ground states occurs at
$T_{\rm K} / \Delta \simeq 0.3$.
In the NRG calculation, the lowest-lying $\sim$500 states are kept at each renormalization
step.
}
\label{fig:2}
\end{figure}
\subsection{Multiorbital scattering impurity}
Next we discuss a case of multiorbital scattering impurity to distinguish the
two types of superconductivity in the Kondo effect.
For this purpose, we have derived the NRG Hamiltonian (\ref{eqn:2.24}).
For $\tilde{\Delta} = 0$ and $\tilde{J}_0 = \tilde{J}_1$ ($\tilde{J}_2$),
the model corresponds to the two-channel Kondo model.
\cite{Nozieres}
In this metallic case, the Kondo temperature is given by
$T_{\rm K} = |J_0 \rho| \exp (-1 / |J_0 \rho|)$.
In the logarithmic discretization procedure for NRG, $J_0 \rho$ is effectively reduced
by the factor $g(\Lambda)$:
\cite{Satori}
$g(3) = 0.910$ is taken here.
\par

\subsubsection{$S_{\rm imp} = 1/2$ impurity case}
Let us begin with the $S_{\rm imp} = 1/2$ impurity.
For the $p_x + i p_y$-wave, the ground and the first excited states are always
spin-doublet ($S = 1/2$) and particle-hole doublet ($S = 0$), respectively, which
is same as the $\tilde{J}_1 = 0$ case discussed above.
The $T_{\rm K} / \Delta$ dependence of the first excited state energy is also similar,
but it does not approach zero for large values of $T_{\rm K} / \Delta$.
Figure~\ref{fig:1} shows the ${\tilde J}_1$ dependence of the first and second excitation
energies when ${\tilde J}_0$ and ${\tilde \Delta}$ are fixed at $0.2$ and $10^{-5}$,
respectively, which give $T_{\rm K} / \Delta = 13.8$ for $\tilde{J}_1 = 0$.
The first excited energy $(E_{\rm B,1} / \Delta)^*$ given in Fig.~\ref{fig:1}
increases gradually with increasing $|\tilde{J}_1|$ for $|\tilde{J}_1| < 0.2$.
This means that the effect of $p$-wave scattering does not destabilize the $\tilde{J}_1=0$
state immediately.
The second excitation energy level $(E_{\rm B,2} / \Delta)^*$ is located above 1.0.
When ${\tilde J}_1$ is close to ${\tilde J}_0$,
$(E_{\rm B,1} / \Delta)^*$ upturns suddenly and
$(E_{\rm B,2} / \Delta)^*$ appears below 1.0 as the second bound state.
Both energy levels cross just when ${\tilde J}_1$ is equal to ${\tilde J}_0$.
When ${\tilde J}_0$ and ${\tilde J}_1$ are exchanged, the same result is obtained.
For ${\tilde \Delta} = 0$, we obtain the non-Fermi liquid fixed point derived from the
two-channel Kondo model
\cite{Nozieres}
for $\tilde{J}_0=\tilde{J}_1$.
Once the two couplings are anisotropic in this case,
the non-Fermi liquid fixed point becomes unstable immediately,
and a Fermi-liquid fixed point is stabilized by a single relevant channel.
When $\Delta$ is finite for $\tilde{J}_1=\tilde{J}_0$,
the two independent channels are combined by $\Delta$.
The continuous change of the curve $(E_{\rm B,1}/\Delta)^*$ in Fig.~\ref{fig:1} means that
there is no discontinuous change of the low-energy spectrum
in all the parameter space of $\tilde{J}_0$ and $\tilde{J}_1$.
The abrupt change of $(E_{\rm B,1}/\Delta)^*$ around $\tilde{J}_1=\tilde{J}_0$
corresponds to the crossover from the $\tilde{J}_0$ dominant spectrum
to the $\tilde{J}_1$ dominant one.
\par

\begin{figure}
\begin{center}
\epsfxsize=7.5cm
\epsfbox{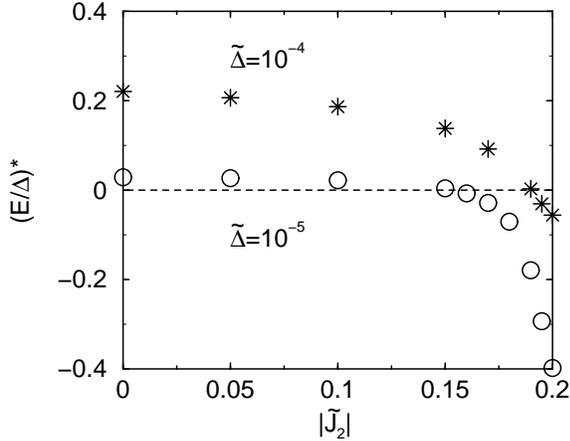}
\end{center}
\caption{
$\tilde{J}_2$ ($<0$) dependence of the energy of the lowest-lying particle-hole doublet
($S = 0$) for $S_{\rm imp} = 1/2$ in the $d_{x^2 - y^2} + i d_{xy}$-wave supercinducting state,
where $\tilde{J}_0$ is fixed at $-0.2$ ($\tilde{\Delta} = 1.5 \Delta$).
The energies are measured from those of the spin doublet ($S = 1/2$).
In the NRG calculation, the lowest-lying $\sim$500 states are kept at each renormalization
step.
}
\label{fig:3}
\end{figure}
Contrary to the above $p_x + i p_y$-wave case, the particle-hole doublet ($S = 0$) can be
the ground state for the $d_{x^2 - y^2} + i d_{xy}$-wave when $T_{\rm K}$ is large enough.
Figure~\ref{fig:2} shows the $T_{\rm K} / \Delta$ dependence of the particle-hole doublet
($S = 0$) energy $(E / \Delta)^*$ measured from the spin doublet ($S = 1/2$) for
$\tilde{J}_0 = \tilde{J}_2$.
The interchange of the ground states is very similar to a $S_{\rm imp} = 1/2$ impurity
in the $s$-wave superconductor where only $l = 0$ orbital is involved.
We note that the particle-hole doublet ($S = 0$) ground state is not smoothly connected to
the $\Delta =0$ limit where a spin-doublet ground state is realized.
\cite{Cragg,Pang}
As $|\tilde{J}_2|$ is decreased from $|\tilde{J}_2| = |\tilde{J}_0|$, the spin-doublet
state is more stable as shown in Fig.~\ref{fig:3}.
In other words, the coupling $\tilde{J}_2$ lowers the energy of the Kondo singlet formed
against the superconducting condensation energy.
Thus the role of the coupling $\tilde{J}_2$ for the $d_{x^2 - y^2} + i d_{xy}$-wave is
completely different from that of the coupling $\tilde{J}_1$ for the $p_x + i p_y$-wave
when their magnitude is close to $|\tilde{J}_0|$.
In Fig.~\ref{fig:1}, the energy of the first excited state ($S = 0$) increases
as $|\tilde{J}_1|$ ($< |\tilde{J}_0|$) becomes larger, implying that the Kondo singlet
becomes unstable.
We conclude that the roles of $l \ne 0$ exchange couplings for the spin-singlet and
spin-triplet superconductors are opposite to each other for $S_{\rm imp} = 1/2$.
\par

\begin{figure}
\begin{center}
\epsfxsize=7.5cm
\epsfbox{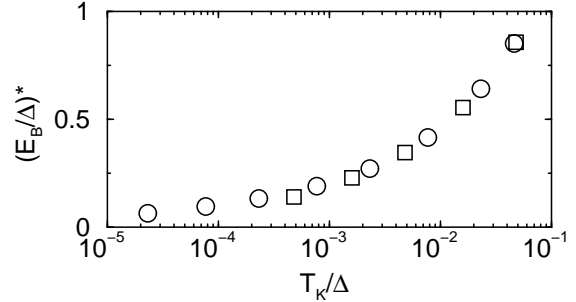}
\end{center}
\caption{
$T_{\rm K} / \Delta$ dependence of the bound state energy for $S_{\rm imp} = 1$
in the $p_x + i p_y$-wave supercinducting state.
The energies are measured from those of the spin singlet ground state ($S_z = 0$).
The bound state is a spin doublet ($S_z = \pm 1$).
The circles and squares represent the data for $\tilde{J}_0 = \tilde{J}_1 = -0.15$ and
for $\tilde{J}_0 = \tilde{J}_1 = -0.2$, respectively.
In the NRG calculation, the lowest-lying $\sim$800 states are kept at each renormalization
step.
}
\label{fig:4}
\end{figure}
\subsubsection{$S_{\rm imp} = 1$ impurity case}
When $T_{\rm K} / \Delta \ll 1$ is satisfied for $S_{\rm imp} = 1/2$, the spin-doublet
is the ground state in both superconducting states.
We can see a more drastic difference between them for the $S_{\rm imp} = 1$ case discussed
below.
The exchange couplings of $l = 0$ and $l = 1$ electrons with the local spin breaks the
spherical symmetry of the total spin in the $p_x + i p_y$-wave superconducting state, and
they lift the degeneracy of spin triplets.
The anisotropy is uniaxial and a spin singlet is stabilized against a doublet in this case.
As a consequence, a singlet ground state is realized even for $T_{\rm K} / \Delta \ll 1$.
Figure~\ref{fig:4} shows the $T_{\rm K} / \Delta$ dependence of the energy of the first
excited doublet state ($S_z = \pm 1$) measured from the singlet ground state ($S_z = 0$)
for $\tilde{J}_1 = \tilde{J}_0$.
This singlet for $T_{\rm K} / \Delta \ll 1$ is not related to a Kondo singlet but to
the spin anisotropy introduced by the exchange coupling $\tilde{J}_1$ in addition to
$\tilde{J}_0$.
When $T_{\rm K} / \Delta$ is large enough, there is no impurity bound state below the
energy gap.
In this sense, the local spin is compensated by the Kondo effect.
The ground state is always a spin singlet except for either $\tilde{J}_0 =0$ or
$\tilde{J}_1 = 0$ case where it is a spin triplet over all the values
of $T_{\rm K} / \Delta$.
\par

\begin{figure}
\begin{center}
\epsfxsize=7.5cm
\epsfbox{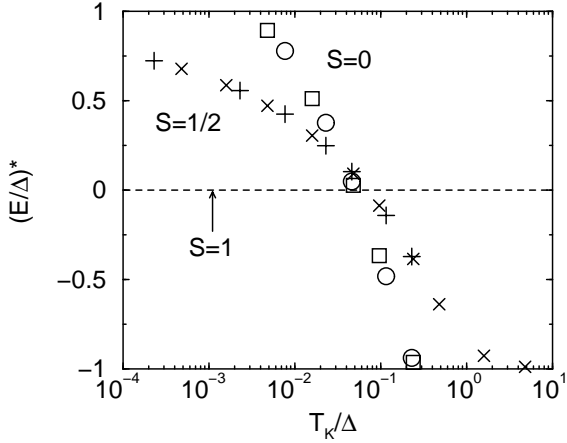}
\end{center}
\caption{
$T_{\rm K} / \Delta$ dependence of the singlet-state ($S = 0$ state) and quartet-state
(twofold degenerate $S = 1/2$ state) energies for $S_{\rm imp} = 1$
in the $d_{x^2 - y^2} + i d_{xy}$-wave supercinducting state.
The energies are measured from those of the spin triplet ($S = 1$).
The circles ($\tilde{J}_0 = \tilde{J}_2 = -0.15$) and squares
($\tilde{J}_0 = \tilde{J}_2 = -0.2$) represent the energies of the singlet ($S = 0$).
The plus ($\tilde{J}_0 = \tilde{J}_2 = -0.15$) and cross
($\tilde{J}_0 = \tilde{J}_2 = -0.2$) symbols represent those of the quartet
($S = 1/2$).
The interchange of the $S = 1$ and $S = 0$ ground states occurs at
$T_{\rm K} / \Delta \simeq 0.05$.
In the NRG calculation, the lowest-lying $\sim$800 states are kept at each renormalization
step.
}
\label{fig:5}
\end{figure}

\begin{figure}
\begin{center}
\epsfxsize=7.5cm
\epsfbox{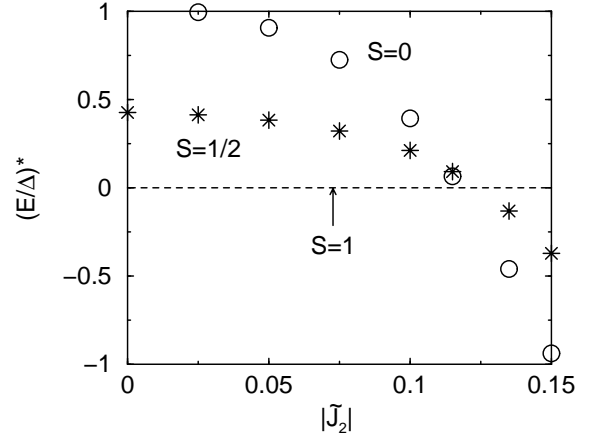}
\end{center}
\caption{
$\tilde{J}_2$ ($<0$) dependence of the singlet-state ($S = 0$ state) and quartet-state
(twofold degenerate $S = 1/2$ state)
energies for $S_{\rm imp} = 1$ in the $d_{x^2 - y^2} + i d_{xy}$-wave supercinducting state.
The fixed parameters are $\tilde{J}_0 = -0.15$ and $\tilde{\Delta} = 10^{-5}$
($\tilde{\Delta} = 1.5 \Delta$).
The energies are measured from those of the spin triplet ($S = 1$).
The circles and stars represent the data for the singlet ($S = 0$) and quartet ($S = 1/2$)
states, respectively.
In the NRG calculation, the lowest-lying $\sim$800 states are kept at each renormalization
step.
}
\label{fig:6}
\end{figure}

On the other hand, for the $S_{\rm imp} = 1$ impurity in
the $d_{x^2 - y^2} + i d_{xy}$-wave case, we find the interchange of spin-triplet
and spin-singlet ground states as found for $S_{\rm imp} = 1/2$.
An energy level of a quartet state (degenerate $S = 1/2$ doublet states) appears between
those of the singlet and triplet states as shown in Fig.~\ref{fig:5}.
The quartet ($S = 1/2$) cannot be the ground state for all the values
of $T_{\rm K} / \Delta$.
As $|\tilde{J}_2|$ is decreased from $|\tilde{J}_0|$, the spin-triplet state is stabilized
as the spin-doublet is stabilized for $S_{\rm imp} = 1/2$.
This is shown in Fig.~\ref{fig:6}.
Contrary to the $S_{\rm imp} = 1/2$ case, the difference between the two superconducting
states is remarkable for $T_{\rm K} / \Delta \ll 1$.
This is due to the spin anisotropy introduced by the two-orbital couplings of the Kondo
effect in the $p_x + i p_y$-wave superconducting state, while the spin of the total system
is isotropic in the $d_{x^2 - y^2} + i d_{xy}$-wave.
\par

\subsection{Impurity susceptibility}
Finally we mention the different low-temperature behavior between two types of magnetic
susceptibility, $\chi_{\rm imp}$ and $\chi_{\rm loc}$ in the above cases.
The former is the impurity susceptibility defined by
$\chi_{\rm imp} = (\chi - \chi_0)$ where $\chi$ ($\chi_0$) is the magnetic susceptibility
of the conduction electron system including (in the absence of) the local spin.
The latter is the local spin susceptibility defined by
$\lim_{h \rightarrow 0} (M_z / h)$ where $M_z = g \mu_{\rm B} <S_z>$ is the
magnetization of the local spin and $h$ is a magnetic field coupled to the local spin.
For $S_{\rm imp} = 1/2$ in the $p_x + i p_y$-wave, the ground state is a spin doublet
for all the values of $T_{\rm K} / \Delta$, as we have discussed.
In this case, both $\chi_{\rm imp}$ and $\chi_{\rm loc}$ shows the Curie law.
In zero temperature limit ($T \rightarrow 0$), $T \chi_{\rm loc}$ decreases
with the increase of $T_{\rm K} / \Delta$, while $T \chi_{\rm imp}$ always reaches
a constant $S_{\rm imp} (S_{\rm imp} + 1)/3$.
Such difference indicates that the local spin is shrunk as the Kondo
effect is stronger and the Cooper pairs couple with the local spin.
On the other hand, $\chi_{\rm loc} (T \rightarrow 0)$ shows a constant value and
$\chi_{\rm imp} (T \rightarrow 0)$ vanishes when a spin-singlet ground state is
realized for $S_{\rm imp} = 1$ in the $p_x + i p_y$-wave superconducting state
and for small values of $\Delta$ in the $d_{x^2 - y^2} + i d_{xy}$-wave.
For the $p_x + i p_y$-wave, $\chi_{\rm loc}$ decreases with the increase of
$\Delta$, implying the Van Vleck like behavior.
\par

\section{conclusion}
We have studied a $S_{\rm imp} = 1/2$ or $S_{\rm imp} = 1$ impurity coupled to
the $p_x + i p_y$-wave
or $d_{x^2 - y^2} + i d_{xy}$-wave superconducting state in order to find the different
behavior between the spin-triplet and spin-singlet Cooper pairs in the Kondo effect.
The Kondo effect in both types of superconductors can be described by the
same model for a short-range scattering impurity with only $l = 0$ electrons.
In this case, only one of the paired electrons couples to the local spin directly.
It means that this type of effective exchange interaction does not depend on
the spin of the Cooper pair.
To see spin effects of Cooper pairs, it is necessary to add the $l \ne 0$ orbital
of the paired electrons to the effective exchange interaction.
We consider $l = 1$ for the $p_x + i p_y$-wave and $l = 2$ for the
$d_{x^2 - y^2} + i d_{xy}$-wave.
The two-orbital exchange interaction generates uniaxial spin anisotropy for
the $p_x + i p_y$-wave since the spin of the Cooper pair is equal to one
and is aligned to the $x$-$y$ basal plain.
On the other hand, spin isotropy is conserved for the $d_{x^2 - y^2} + i d_{xy}$-wave,
since the singlet Cooper pair has no spin.
The two types of Cooper pairs cause the different Kondo behavior as follows:
\par
\noindent
(1)~$S_{\rm imp} = 1/2$ local spin
\par
The ground state is a spin doublet in all the parameter region of $J_0$($<0$),
$J_1$($<0$), and $\Delta$ for the $p_x + i p_y$-wave.
For the $d_{x^2 - y^2} + i d_{xy}$-wave, the interchange of the spin-doublet
(for large $\Delta$) and spin-singlet (for small $\Delta$) ground states is found.
\par
\noindent
(2)~$S_{\rm imp} = 1$ local spin
\par
The ground state is a singlet represented by $S_z = 0$ in all the parameter region
as mentioned above for the $p_x + i p_y$-wave.
It is split from a spin triplet due to the uniaxial spin anisotropy.
For the $d_{x^2 - y^2} + i d_{xy}$-wave, the ground state interchanges between
the spin-triplet (for large $\Delta$) and spin-singlet (for small $\Delta$) states.
\par

We have found that physical quantities such as bound state energies are scaled by
$T_{\rm K} / \Delta$, which is a common feature for such competition between the Kondo effect
and the energy gap.
\par

In order to apply our Kondo model to real systems, we must start from an extended
Anderson model with orbital degeneracy.
If we identify the orbital denoted by $l$ with one of the atomic orbitals of
the impurity, we have to take $l = -1$ ($l = -2$) into account as well as
$l = 1$ ($l = 2$) for the $p_x + i p_y$-wave ($d_{x^2 - y^2} + i d_{xy}$-wave).
Since the $p_x + i p_y$-wave superconducting order parameter makes $l = -1$ couple
with $l = 2$, we need to include the four orbitals ($l = -1, 0, 1, 2$) of conduction
electrons in the present model.
In the same manner, four orbitals ($l = -2, 0, 2, 4$) of electrons are necessary
for the $d_{x^2 - y^2} + i d_{xy}$-wave case.
As for another point, the $S_{\rm imp} = 1$ impurity is actually obtained for
the strong Hund coupling limit.
If the Hund coupling is weak, effective orbital exchange interactions should be present.
Since the conduction-electron part including the BCS interaction term
is transformed to the hopping type of NRG Hamiltonian in Eq.~(\ref{eqn:2.18}) first,
the orbital exchange interactions break the conservation of the total number of
quasiparticles and the $z$ component of the total spin.
This makes the mumerical calculation much harder.
Thus the model becomes more complicated for the orbitally degenerate impurity.
Neverthless, the present form of our model can describe a local moment
in superconducting states if each orbital can be regarded as an independent branch of
a conduction band instead of an atomic orbital.
It is interesting to search for a realistic case where such extention of our study is
possible.

\acknowledgements
This work is supported by JSPS for Encouragement of Young Scientists (No.~13740214).

\end{document}